\documentclass[epj]{svjour}

\usepackage{graphicx}
\usepackage{fancyhdr}
\def\makeheadbox{{
 }}
\def\mytitle{My title} 
\def\myauthors{My name}  
\def\mytype{My type of session}
\def\mysession{My session}

\def\mytitle{Search for Chargino-Neutralino Pair Production at CDF} 
\def\myauthors{Else Lytken}    
\def\mytype{Contributed Talk}    
\def\mysession{Colliders - SUSY Phenomenology}

\newcommand{\met}{$\not \! \! E_T$}
\newcommand{\mass}{GeV/$c^2$}
\newcommand{\pt}{p$_T$}

\begin{document}
\title{Search for Chargino-Neutralino Pair Production at CDF}
\author{Else Lytken \newline for the CDF Collaboration
}                     
%
%
\institute{Purdue University, Lafayette IN. (Now at CERN). }
%
\date{}
\abstract{
We present the results of a search for associated production of the lightest 
chargino and next-to-lightest neutralino using 1 fb$^{-1}$ of $\sqrt s$  = 1.96 TeV p$\rm \bar p$
data collected with the CDF detector at the  Tevatron. We combine the results of 
several multi-lepton final states  to set upper limits on the cross section 
times branching ratio for  chargino-neutralino production as a function of the 
chargino mass.
\PACS{
    {}{}   
     } 
} 
\maketitle
\section{Introduction}
\label{intro}
The associated production of $\tilde \chi_1^\pm$ and $\tilde \chi_2^0$ is regarded as the golden discovery channel at the Tevatron for Supersymmetry (SUSY). Compared to the expected production cross sections of other SUSY processes for non-excluded masses, the chargino-neutralino pair production dominated by W-exchange is favorably high. In addition most of the allowed parameter space has a significant branching ratio into leptons. Typically a pair of opposite-sign leptons is expected to come from the neutralino decay, $\tilde  \chi_2^0 \rightarrow \ell^\pm \ell^\mp \tilde \chi_1^0 $, via a virtual Z or a slepton. The main decay mode of low mass charginos is into a charged lepton, a neutrino, and a $\tilde \chi_1^0$. This gives a clean and striking signature of three leptons, and missing transverse energy, $\not \!  \! \! E_T$, from the remaining final state particles; a neutrino and two $\tilde \chi_1^0$'s. Assuming R-parity\footnote{$R_{\rm P}=(-1)^{3(B-L)+2s}$} is conserved the two lightest neutralinos will escape undetected. \\
Previously published searches have set limits on the chargino mass of 103.5 \mass~\cite{Ref1} overall, and 117 \mass\ for a particular model~\cite{Ref2}.
\section{Analysis and prerequisites}
\label{sec:1} 
Despite the low standard model backgrounds, the search suffers from a small total cross section times branching fraction into three charged leptons ( $<$1 pb). A significant fraction of the leptonic decays are also expected to go into 
$\tau$'s, resulting in low \pt\ or non-leptonic final states. This analysis searches for three isolated charged leptons, electrons or muons only, and several dedicated approaches are combined to obtain maximum sensitivity. Three analyses uses high \pt\ (20 GeV/c) single lepton triggers: electron + 2$e/\mu$, ($e\ell\ell$), muon + 2 $e/\mu$, ($\mu\ell\ell$), and electron or muon with same-sign electron or muon ($e^\pm e^\pm$,$e^\pm\mu^\pm$,$\mu^\pm\mu^\pm$). In these cases we can relax the requirements on the second and third leptons (if applicable) to improve the acceptance. To cover regions in parameter space with lower \pt\ leptons we also have two analyses taking advantage of low \pt\ (4 GeV/c) dilepton triggers: $\mu\mu$ + $e/\mu$, ($\mu\mu\ell$), and ee + isolated track ($eet$). In the latter case we are also sensitive to some hadronically decaying $\tau$'s.\\
The data are collected by the CDF II detector~\cite{Ref5} between Spring 2002 and Spring 2006. This corresponds to an integrated luminosity of ~0.7 fb$^{-1}$ for the $\mu\ell\ell$ selection and ~1 fb$^{-1}$ for the others. The analyses described here are published in~\cite{Ref3}. \\

\subsection{Analysis cuts}
Once we require three leptons the background is already greatly reduced. Remaining background processes are $Z$+$\gamma$, where the photon converts; $Z$+jets, where the jet is misidentified as a lepton; heavy-flavor background ($t\bar t$, $b\bar b$), and signal-like background from $WZ$, $ZZ$. For the same-sign dilepton analysis, the main backgrounds are $W$+$\gamma$, $W$+jets, $Z$+$\gamma$, and $WW$. We estimate the background from misidentified leptons from jet-triggered data samples, and other backgrounds from simulation. In the case of $\mu\mu\ell$, the heavy-flavor background is also estimated from data.\\
Detailed knowledge of the lepton identification criteria and photon conversion tagging is the best handle to reduce backgrounds without real, isolated, and prompt leptons. In addition, we only accept events with low jet activity, either by direct requirements on the number of reconstructed jets, or by constraining the $\Sigma E_T$ of events.
We suppress $Z$+$\gamma$ by asking for a minimal missing transverse energy, \met $\geq $ 15 (or 20) GeV. 
To reduce $WZ$, $ZZ$, and further suppress $Z$+$\gamma$, we also require that the invariant mass of opposite-sign leptons must not fall in the range 76-106 \mass, nor be less than 15 \mass\ (20 \mass\ for some analyses).

\subsubsection{Control samples} 
\begin{figure}[h]
\begin{center}
\includegraphics[width=0.45\textwidth]{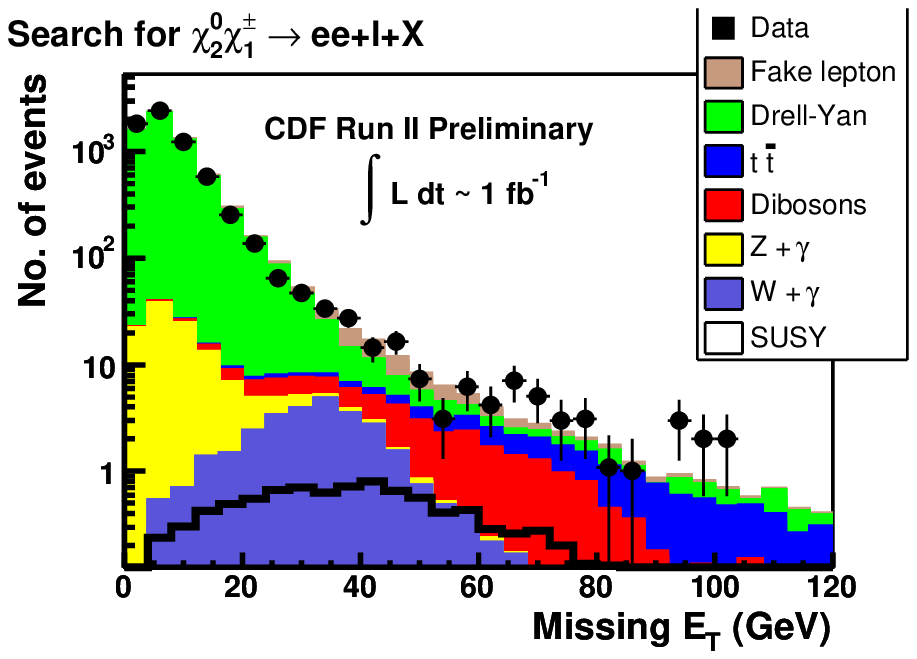}
\includegraphics[width=0.43\textwidth,height=0.25\textheight]{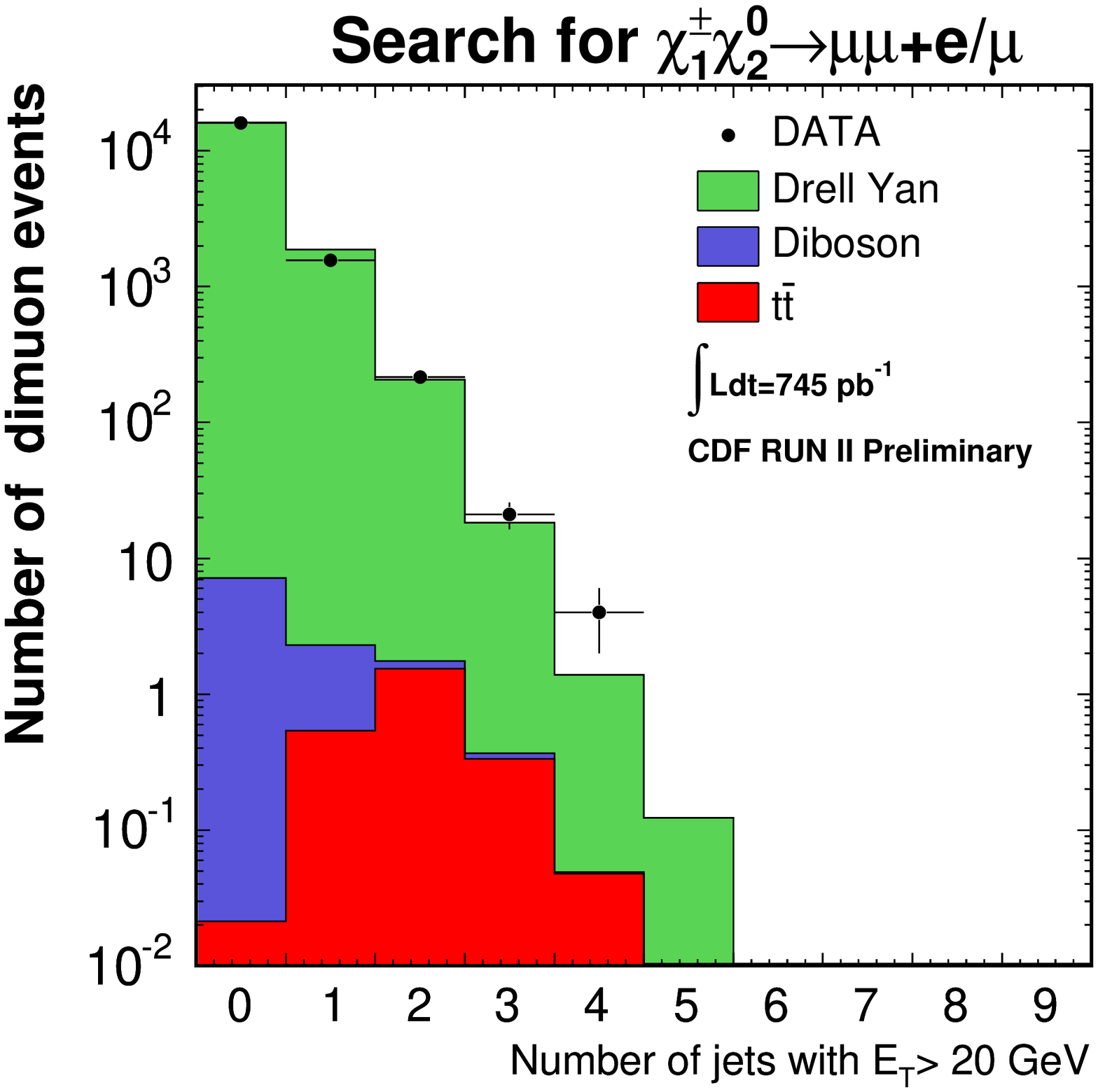}
\includegraphics[width=0.45\textwidth]{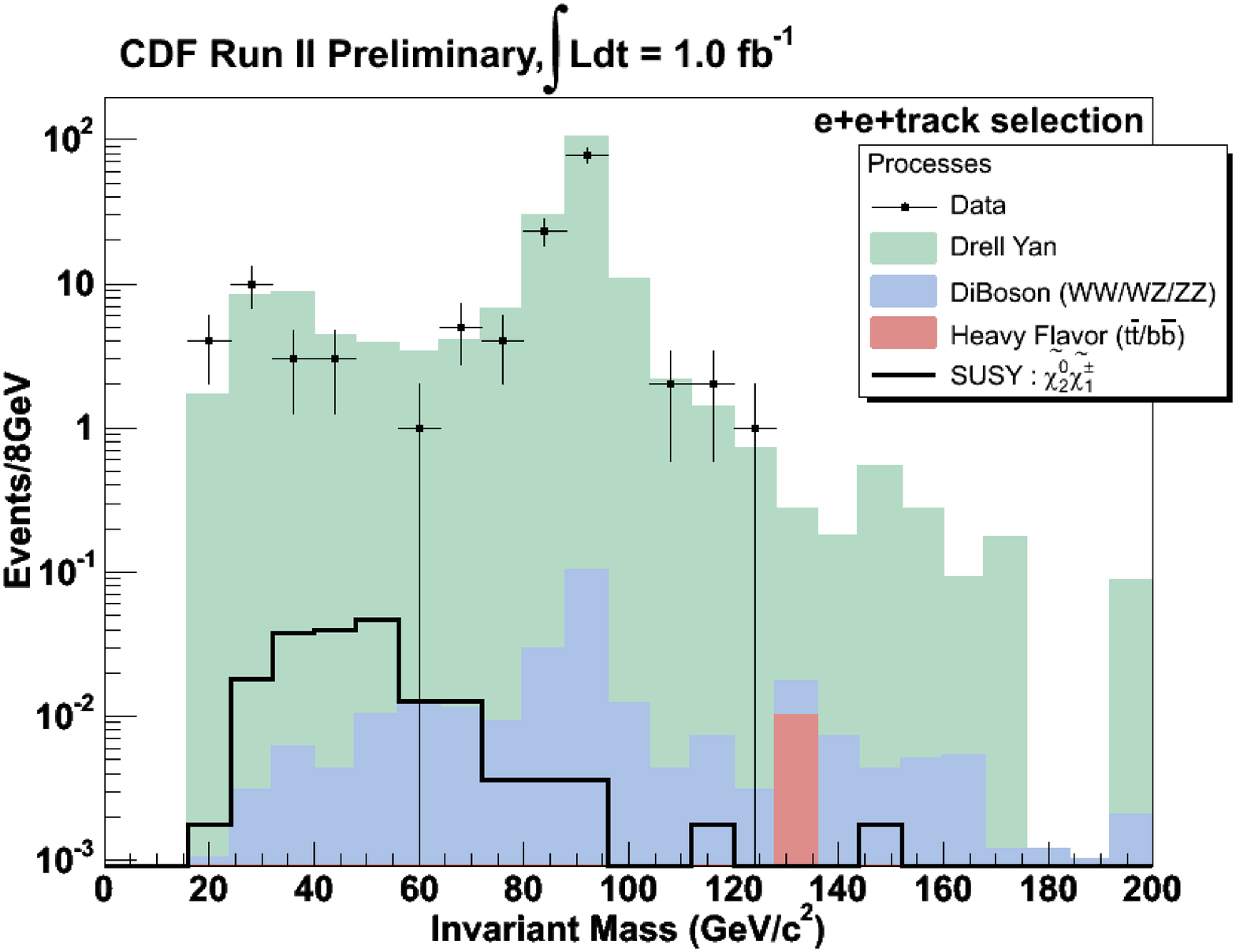}
\caption{Examples of control samples. From top: \met\ in ee events, number of jets with $E_T \geq $ 20 GeV in $\mu\mu$ events, and the invariant mass of opposite-sign dielectrons in events with 2 electrons and an isolated track. Error bars are statistical.}
\label{fig:1}       
\end{center}
\end{figure}

Before we look for a possible excess of events passing the analysis cuts, we test our background predictions extensively by comparing them to observations in control samples, where we do not expect significant contributions from new physics. We check both the total event counts and the shapes of the analysis variables. A few examples can be found in Figure~\ref{fig:1}.
The benchmark point used for illustration in the plots is an mSUGRA point with m$_0$ = 100 \mass, m$_{1/2}$ = 180 \mass, tan$\beta$ = 5, A$_0$ = 0, and $\mu >$0, with a chargino mass of 113 \mass, and $\sigma\times Br$ = 0.16 pb.\\
We estimate systematic uncertainties from the identification of leptons, the integrated luminosity, initial- and final state radiation, parton density functions, the jet energy scale, the estimate of jets or photon conversions misidentified as prompt leptons, and theoretical uncertainties on the cross sections. Typically the largest systematical uncertainties come from the estimation of misidentified lepton. On average the systematical uncertainty estimate is about 15\% and 20\% for signal and background, respectively. 

\section{Results}
\label{sec:2}
After verifying that we have good agreement between expectation and observation in our control samples we proceed to look at the subset of events passing the analysis requirements. The results are shown in Tables~\ref{tab:1} and~\ref{tab:2}. For completeness we show also the expected yield from the mSUGRA benchmark point described above. 
%
\begin{table}
\begin{center}
\caption{Results for trilepton channels}
\label{tab:1}       
\begin{tabular}{lccccc}
\hline\noalign{\smallskip}
& $e\ell\ell$ & $\mu\ell\ell$ & $eet$ & $\mu\mu\ell$   \\
\noalign{\smallskip}\hline\noalign{\smallskip}
Background & 0.8$\pm$0.4 & 1.3$\pm$0.3  & 1.0$\pm$0.3 & 0.4$\pm$0.1\\
Exp. signal & 2.1$\pm$0.2 & 2.3$\pm$0.3 & 2.0$\pm$0.1 & 0.6$\pm$0.1\\ \hline
Observed & 0 & 1 & 3 & 1 \\
\noalign{\smallskip}\hline
\end{tabular}\end{center}\end{table}
\begin{table}
\begin{center}
\caption{Results for dilepton channels}
\label{tab:2} 
\begin{tabular}{lccc}
\hline\noalign{\smallskip}
& $e^\pm e^\pm$ & $e^\pm\mu^\pm$ & $\mu^\pm\mu^\pm$  \\
\noalign{\smallskip}\hline\noalign{\smallskip}
Background & 3.0$\pm$0.5 & 4.0$\pm$0.6 & 0.9$\pm$0.1 \\ \hline
Exp. signal & 0.6$\pm$0.1 & 1.7$\pm$0.2 & 1.0$\pm$0.1 \\
Observed & 4 & 8 & 1 \\
\noalign{\smallskip}\hline
\end{tabular}
\end{center}
\end{table}
There are small excesses in the observed event count for some channels but no significant excesses from the expected background are seen. Therefore we use these results to set limits on the chargino-neutralino cross section times their branching ratio into leptons (including $\tau$'s).

\subsection{Interpretation}
\begin{figure}
\begin{center}
\includegraphics[width=0.5\textwidth,height=0.45\textwidth]{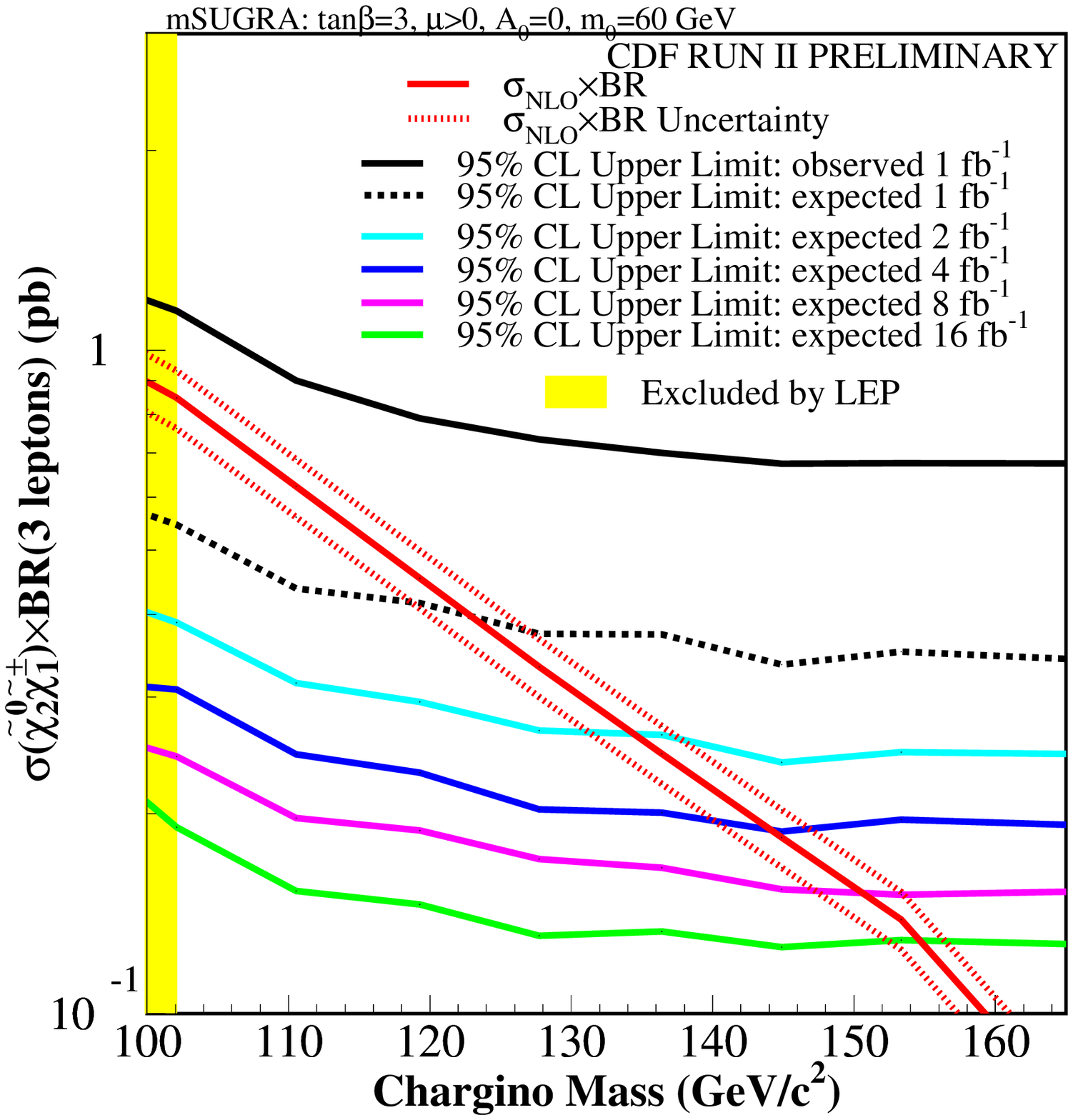}
\includegraphics[width=0.5\textwidth,height=0.45\textwidth]{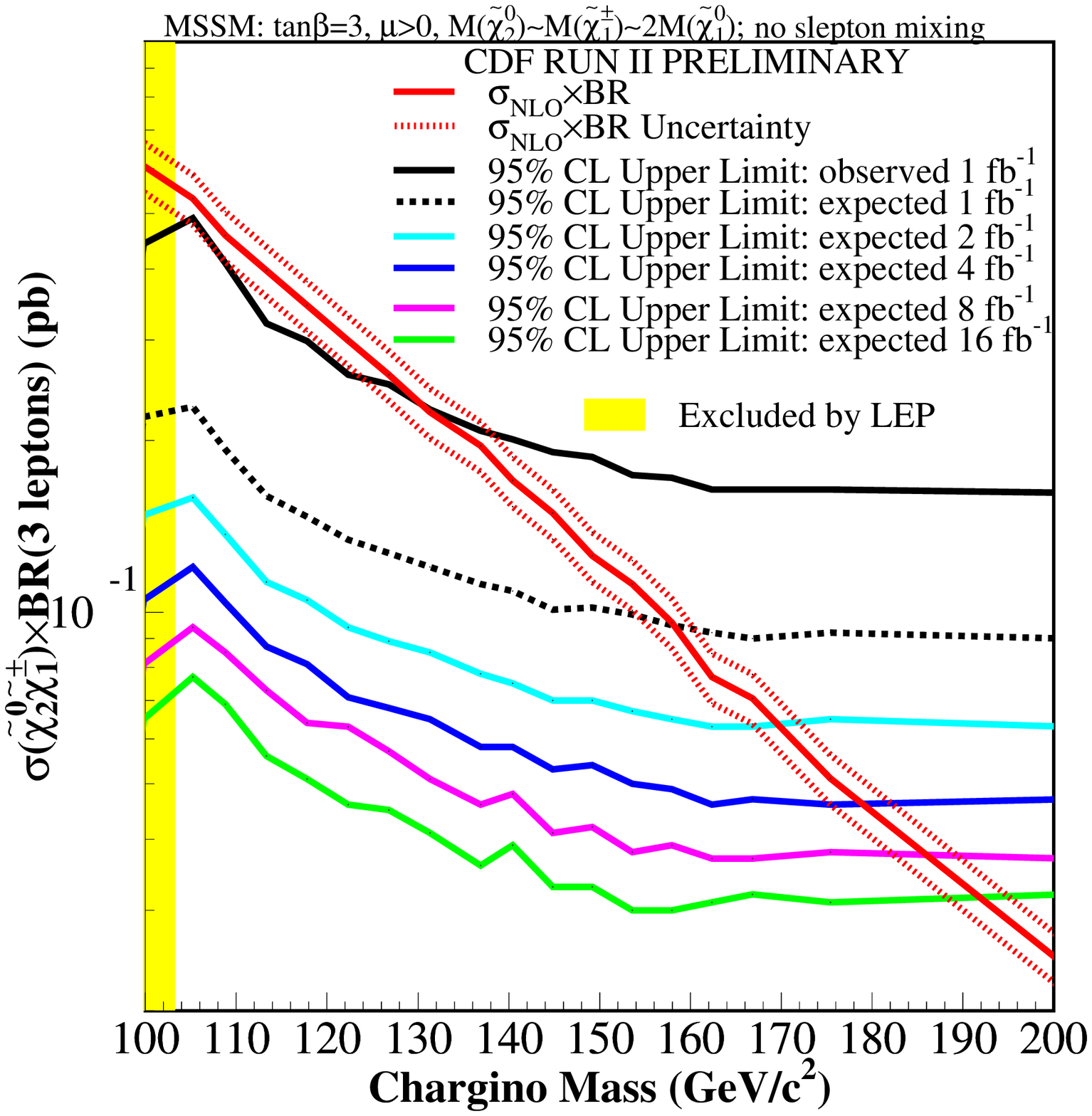}
\includegraphics[width=0.5\textwidth,height=0.45\textwidth]{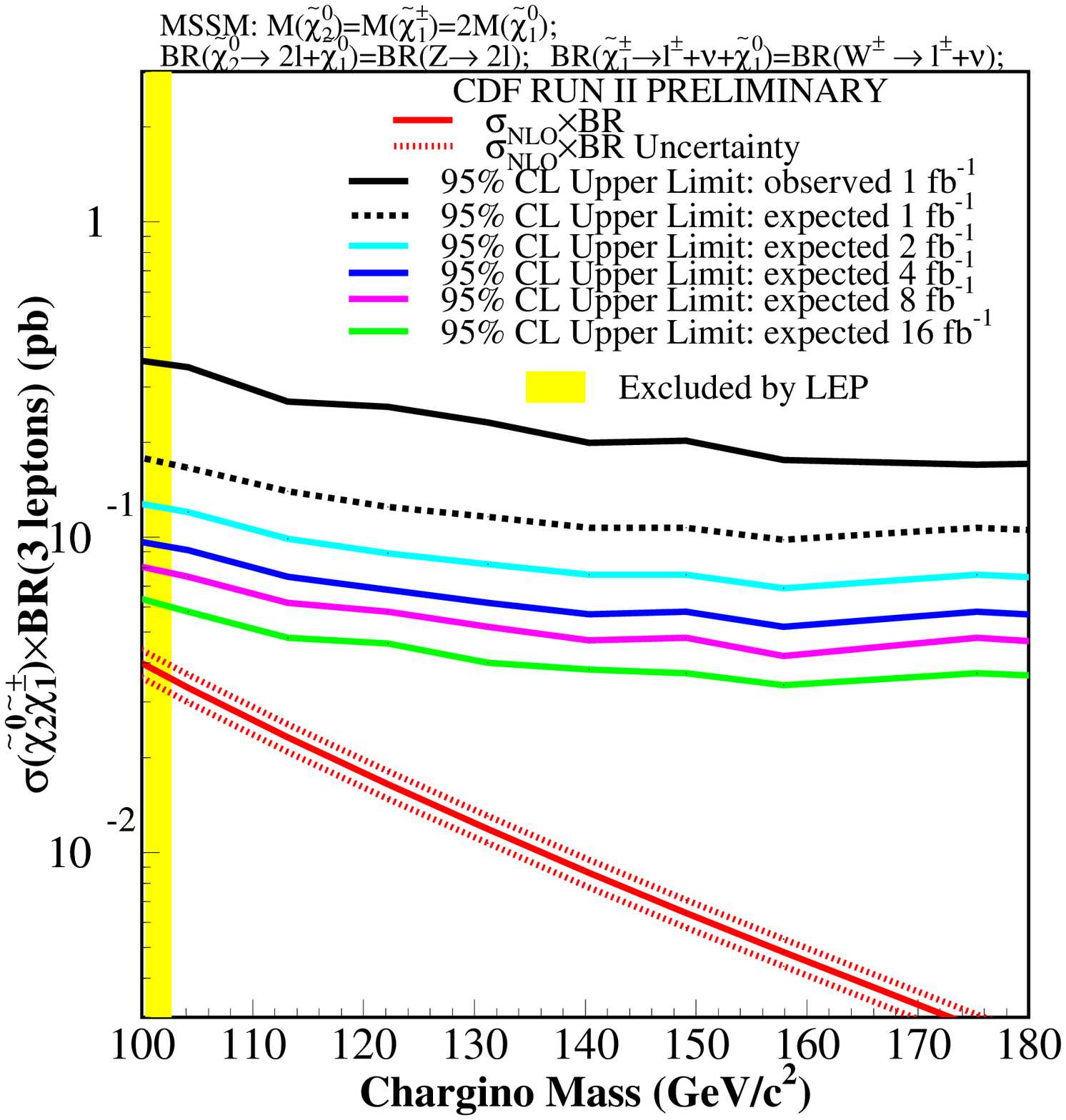}
\caption{Projections to higher luminosities, and the current expected and observed limits for the 3 models. The black curves are the current results, extrapolated to higher masses.}
\label{fig:2}       
\end{center}
\end{figure}

We choose to interpret the result as limits on $\sigma \times Br$ as a function of the chargino mass in 3 models. The first one is a standard mSUGRA scenario with m$_0$ = 60 \mass, tan$\beta$ = 3, A$_0$ = 0, $\mu >$ 0, and m$_{1/2}$ in the range 162-230 \mass. This was found to be the area of parameter space where the analyses had best sensitivity. The second model is similar to the previous one, but with the hypothesis of no slepton mixing and degenerate slepton masses. To keep the same decay modes, we also changed the m$_0$ value to m$_0$ = 70 \mass. Our third scenario is also mSUGRA inspired, but we fix the branching ratios of the $\tilde \chi_1^\pm$ and $\tilde \chi_2^0$ to be equivalent to the low leptonic branching ratios of standard model W's and Z's: BR($\tilde \chi_1^+ \rightarrow \ell \nu \tilde \chi_1^0$) = BR(W$\rightarrow\ell\nu$), and BR($\tilde\chi_2^0\rightarrow\ell\ell\tilde\chi_1^0 $) = BR(Z$\rightarrow\ell\ell$). \\
We present the results in Figure~\ref{fig:2} as 95\% confidence limits using a frequentist approach~\cite{Ref4} that takes into account the correlations between the uncertainties and between channels. The expected number of events in Table~\ref{tab:1} includes events shared among the channels and when calculating the combined acceptance, this overlap is removed, and the background rescaled accordingly. To extract expected and observed mass limits, we include the theoretical uncertainty (represented in red dashed lines in Figure~\ref{fig:2}) in the limit calculation and take the intersection between this and the central value of the theory curve. The expected mass limit in the mSUGRA case was 122 \mass, whereas the observed is below the LEP limit. For the no slepton mixing scenario we set a mass limit of 129 \mass, expecting 157 \mass. We do not yet have sensitivity to the model with reduced lepton decays. The difference between the expected and observed limits corresponds to about 2$\sigma$ and is caused mainly by the excesses observed in the $eet$ and $e^\pm \ell^\pm$ channels. \\
In addition, Figure~\ref{fig:2} shows basic projected limits from these results, as expected with higher integrated luminosities. The curves are extrapolated beyond chargino masses of $\sim$150 \mass\ for mSUGRA,  $\sim$160 \mass\ for the W/Z decay model, and  $\sim$170 \mass\ for the scenario with no slepton mixing. It is also assumed that the uncertainties scale with the luminosity and that no improvements are made to the analyses. 

\section{Conclusion}
We have searched for pair production of charginos and neutralinos in the CDF Run II dataset corresponding to between 0.7 and 1 fb$^{-1}$. No significant excess with respect to the expectation from the standard model was observed. We show exclusion limits on the production cross section times branching ratio as a function of the chargino mass, and a simple projection to larger datasets. In one mSUGRA inspired model with no slepton mixing and degenerate slepton masses, we can exclude chargino masses below 151 \mass\ with 95\% confidence with the first fb$^{-1}$.

%
%

\end{document}